\documentclass[10pt,conference]{IEEEtran}

\usepackage{algorithmic,amsmath,amssymb,amsfonts,booktabs,braket,cite,graphicx,hyperref,mathbbol,pgfplots,quantikz,ragged2e,stfloats,textcomp,tikz}
\usepackage[dvipsnames]{xcolor}
\tikzset{>=latex}
\pgfplotsset{compat=newest}

\hypersetup{
  colorlinks= true,
  urlcolor = blue,
  linkcolor = blue,
  citecolor = blue
}

\DeclareMathOperator{\spn}{span}

\tikzstyle{block}=[draw,minimum size=2em]
\tikzstyle{l}=[minimum size=1.5em]
\tikzstyle{xor}=[draw, minimum size=0.8em,append after command={[shorten >=\pgflinewidth, shorten <=\pgflinewidth,] 
        (\tikzlastnode.north) edge (\tikzlastnode.south)
        (\tikzlastnode.east) edge (\tikzlastnode.west)
}]
\def\BibTeX{{\rm B\kern-.05em{\sc i\kern-.025em b}\kern-.08em
    T\kern-.1667em\lower.7ex\hbox{E}\kern-.125emX}}

\begin{document}

\title{Simon's Algorithm for the Even-Mansour Cipher\\on Quantum Hardware}

\author{
\IEEEauthorblockN{Anina Köhler\IEEEauthorrefmark{1}, Jakob Murauer\IEEEauthorrefmark{2}, Tim Heine\IEEEauthorrefmark{3}, Stefan Rosemann\IEEEauthorrefmark{4} and Tobias Hemmert\IEEEauthorrefmark{4}}
\IEEEauthorblockA{\IEEEauthorrefmark{1}Department of Computer Science, University of Oxford, Oxford, OX1 3QD, UK}
\IEEEauthorblockA{\IEEEauthorrefmark{2}Research Institute CODE, University of the Bundeswehr Munich, 85579 Neubiberg, Germany}
\IEEEauthorblockA{\IEEEauthorrefmark{3}Institute of Quantum Technologies, German Aerospace Center (DLR), 89081 Ulm, Germany}
\IEEEauthorblockA{\IEEEauthorrefmark{4}Federal Office for Information Security (BSI), 53175 Bonn, Germany}
}

\maketitle

\begin{abstract}
Simon's algorithm is a polynomial period-finding algorithm that has been used to exploit the algebraic structure of specific symmetric ciphers, showing that exponential speedups in their cryptanalysis are theoretically possible.
While the theoretical framework for an attack using Simon's algorithm on the Even-Mansour cipher is well-established, practical implementations on noisy intermediate-scale quantum (NISQ) hardware remain limited.
This paper presents a proof of concept quantum cryptanalysis of the Even-Mansour cipher using Simon’s period-finding algorithm on NISQ hardware. 
For $N=3$ and $N=4$, we successfully demonstrate secret key recovery for $N$-bit constructions on the ibm\_miami processor. 
Our experiments also identify a scaling limitation in the classical pre-processing stage: The DORCIS circuit optimization tool encountered a memory bottleneck at $N=5$, preventing the generation of optimized circuits for larger key lengths. 
Our results suggest firstly that Simon's algorithm is effective for the Even-Mansour cipher for short bit lengths on current quantum hardware. 
Secondly, while DORCIS is effective for the small-scale S-boxes for which it was designed, there remains a need for the investigation of more scalable and efficient synthesis tools capable of handling larger and more general permutations in the context of Even-Mansour ciphers.
\end{abstract}


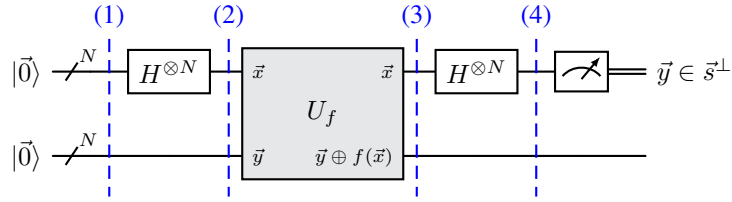
\begin{figure*}[ht!]
\centering
\begin{quantikz}[slice all,remove end slices=1,slice
titles=\textcolor{blue}{(\col)},slice style=blue]
    \lstick{$\ket{\vec 0}$} & \qwbundle{N} & \gate{H^{\otimes N}} & \gate[2, style={fill=Gray!20,inner xsep=2pt}][2cm]{U_f}\gateinput{$\vec x$}\gateoutput{$\vec x$} & \gate{H^{\otimes N}} & \meter{} & \rstick{$\vec y\in\vec s^\bot$} \setwiretype{c} \\
    \lstick{$\ket{\vec 0}$} & \qwbundle{N} & & \gateinput{$\vec y$}\gateoutput{$\vec y\oplus f(\vec x)$} & & & 
\end{quantikz}
    \caption{Schema of the $2N$-qubit quantum circuit for Simon’s algorithm used to solve the hidden period problem of $U_f$.}
    \label{fig:sc}
\end{figure*}

\section{Introduction}
\label{sec:introduction}

Symmetric cryptography is assumed to be relatively resistant to quantum attacks. 
Grover’s algorithm \cite{grover} offers a quadratic speedup over classical algorithms for unstructured search problems.
It can be applied to any problem where we can efficiently verify a solution. Thus, it can accelerate an exhaustive search for cryptographic keys.
It is therefore usually deemed sufficient to double the key length in order to maintain the security of symmetric cryptographic schemes against quantum computers.
However: 
\begin{enumerate}
    \item there is often little certainty regarding the underlying computational problems and their complexity,
    \item the scaling of quantum resources through hardware and software innovations is constantly advancing, and
    \item there is a possibility of (quantum)-algorithmic improvements showing that fewer actual resources are needed.
\end{enumerate}  
Hence, it remains an important problem to better understand what is computable specifically with quantum resources and to derive appropriate protection measures.

Key recovery using quantum period-finding is a well-established approach targeting certain symmetric cryptographic primitives. 
The idea is to exploit structural properties in a cipher to take advantage of an exponential speedup of quantum over classical algorithms by reducing the problem of finding the key of a cipher to the problem of finding the hidden period of a periodic function---a problem that can be solved efficiently.
Simon's polynomial period-finding algorithm has been utilized to target quantum extensions of symmetric ciphers, thus demonstrating that exponential speedups are theoretically possible \cite{kuwakadofeistel,kuwakado,kaplan,santoli,em_hosoyamada}.

However, the approach requires the attacker to query an encryption oracle about quantum states, which means that the encryption function (involving the secret key) is used inside a quantum algorithm. 
While this is a common attack model in quantum cryptanalysis \cite{boneh,kaplan,anand}, the attack is considered impractical.

To reduce the number of superposition queries and extend the application to related symmetric constructions, combinations of Simon's algorithm with Grover's algorithm have been proposed.
The idea is to still leverage periodicity to find the secret key, but instead of directly defining a periodic function, to define a family of functions, one of which is periodic. 
Grover's algorithm is used to search through the family of functions and Simon's algorithm internally identifies the periodic one, essentially offering a tradeoff between the size of the Grover search space and the number of classical queries and superposition queries.
The Grover-meets-Simon algorithm uses superposition queries and has been applied to target the ciphers DESX, PRINCE and PRIDE \cite{leander}. 
The Offline Simon's algorithm has been utilized to target the symmetric block ciphers FX and Even-Mansour \cite{bonnetainoffline}, Chaskey, PRINCE, Elephant \cite{bonnetainpractice}, and 2XOR \cite{bonnetainbeyond} in a weaker attack model where the attacker is limited to classical queries to an encryption oracle.
In particular, the application of the Offline Simon's algorithm to 2XOR for the first time proved a more than quadratic (2.5) speedup compared to a symmetric cipher's best classical attack using only classical queries.
The attacks also reveal that common key length extension techniques can be less effective in a quantum setting.

Using substitution boxes (S-boxes) that we constructed in a similar way to the S-box used in the AES algorithm, we implement a proof of concept of the attack on the minimal symmetric Even-Mansour (EM) cipher using Simon's algorithm \cite{kuwakado} on quantum hardware. 
Since EM is underlying many block cipher constructions, including 1-round AES, and the application of Simon's algorithm is the core idea in all related attacks, we believe that a practical evaluation of concrete examples on current hardware usefully illustrates the concept and can offer valuable insights.

\section[Key Recovery for the Even-Mansour Cipher using Simon's Algorithm]{\texorpdfstring{Key Recovery for the Even-Mansour\\Cipher using Simon's Algorithm}{Key Recovery for the Even-Mansour Cipher using Simon's Algorithm}}
\label{sec:methods}

\subsection{Simon's Algorithm}

Let $f:\{0,1\}^N\rightarrow \{0,1\}^N$ be a function on binary strings that is two-to-one.  
Suppose further that there is a $N$-bit string $\vec s$ such that for all $\vec x\neq\vec y$, the equality $f(\vec x)=f(\vec y)$ holds if and only if $\vec y=\vec x\oplus \vec s$, where $\oplus$ is the bitwise exclusive or. This property is called \textbf{Simon's promise}. 
\textbf{Simon's problem} asks to find that hidden period $\vec s$ of $f$.

If $f$ can be evaluated in superposition using the query gate 
\begin{eqnarray*}
    U_f(\ket{\vec x}\ket{\vec a}) = \ket{\vec x} \ket{\vec a\oplus f(\vec x)},
\end{eqnarray*}
then due to interference on the collisions of $f$, Simon's algorithm \cite{simon94, simon97} can be used to sample $N-1$ linearly independent vectors orthogonal to $\vec s$ uniformly at random. 
These vectors can be used to obtain $\vec s$ in $\mathcal{O}(N)$ in terms of query complexity.
This is providing an exponential advantage over classical query algorithms since the best classical strategy is to randomly query $f$ until a collision has been found. Any classical solution therefore necessarily has a worst case runtime that is exponential in $N$ \cite{simon97}.

For $\vec v\in\{0,1\}^N$, let 
\begin{eqnarray*}
\vec v^\bot=(\spn\{\vec v\})^\bot = \{\vec y \in \{0,1\}^N : \vec y\cdot\vec v = 0\}
\end{eqnarray*}
denote the orthogonal complement of the subspace spanned by $\vec v$ in $\{0,1\}^N$.
The key idea of Simon's algorithm is to notice that $H^{\otimes N}(\ket{\vec x}+\ket{\vec x\oplus\vec v})$, which is the Hadamard transform applied to a superposition of two states $\ket{\vec x}$ and $\ket{\vec x\oplus\vec v}$ that have the same amplitude and differ by $\vec v$, gives
\begin{eqnarray*}
\frac{1}{2^N}\sum_{\vec y\in \vec v^\bot}((-1)^{\vec x\cdot\vec y}+(-1)^{(\vec x\oplus\vec v)\cdot\vec y})\ket{\vec y},
\end{eqnarray*}
which is a superposition of states that are orthogonal to $\vec v$.
This observation is useful because by measuring this state, a vector $\vec y\in \vec v^\bot$ is obtained. By creating interference on the collisions of $f$, Simon's algorithm uses this result to sample vectors orthogonal to $\vec s$ uniformly at random.

The schema of the quantum circuit for Simon's algorithm is illustrated in Fig.~\ref{fig:sc}:
\textcolor{blue}{(1)} The input and output registers are prepared in the all zero state. 
\textcolor{blue}{(2)} A layer of $N$ Hadamard gates creates a uniform superposition of all possible basis states to create interference in the input state.
\textcolor{blue}{(3)} The quantum circuit implementing $U_f$ encodes the problem-specific phase shift by pairing up the collisions.
\textcolor{blue}{(4)} A second layer of $N$ Hadamard gates on the input register transforms the state back to the original basis, making the interference visible.
At this stage, a measurement of the input register in the computational basis yields a random vector $\vec y\in \vec s^\bot$. 

This process is repeatedly run to obtain a sequence of vectors $\vec y_1,\vec y_2,\ldots\in \vec s^\bot$. 
Since $\{0,1\}^N$ with addition modulo $2$ is a vector space over the finite field $\mathbb{F}_2=\mathbb{Z}/2\mathbb{Z}$ and $\spn\{\vec s\}$ is a $1$-dimensional subspace, $\vec s^\bot$ is of dimension $N-1$ and contains exactly half of the vectors. 
Once $N-1$ linearly independent $\vec y_i$ have been obtained, a system of linear equations can be solved to obtain the solution $\vec s$.
If Simon's circuit is run $N+r$ times, then the probability that the correct solution $\vec s$ is obtained from the equations is at least $1-2^{-r}$.
The post-processing step to solve for $\vec s$ can be done efficiently using Gaussian elimination.
In total, Simon's algorithm therefore requires $\mathcal{O}(N)$ iterations, and hence quantum queries, and $\mathcal{O}(N^3)$ classical operations in the post-processing step.

\subsection{Even-Mansour Cipher}

The Even-Mansour scheme \cite{em93, em97} is a simple symmetric block cipher construction based on a single random or pseudorandom public permutation. 
It was introduced as a minimal block cipher which has a formal proof of security.
The scheme xors the plaintext with a prewhitening key before applying the permutation and xors the result with a postwhitening key. 

\begin{figure*}[ht]
\centering
\includegraphics[width=1.0\linewidth]{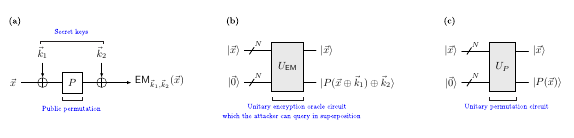}
\caption{Representation of (a) classical Even-Mansour encryption, (b) secret unitary encryption oracle $U_{\sf EM}$, (c) synthesized unitary permutation oracle $U_P$.}
 \label{fig:cryptographic-primitives}
\end{figure*}

Let $P:\{0,1\}^N\to\{0,1\}^N$ be a public random bijective map on $N$-bit strings, which we will subsequently refer to as a permutation on $N$-bit strings, and let $\vec k_1, \vec k_2\in\{0,1\}^N$ be two secret keys chosen uniformly at random.
For a message $\vec x\in\{0,1\}^N$, define the Even-Mansour encryption
\begin{eqnarray*}
\mathsf{EM}_{\vec k_1,\vec k_2}(\vec x)=P(\vec x\oplus \vec k_1)\oplus \vec k_2.
\end{eqnarray*}
The cipher processes data in fixed size blocks of length $N$, while the key $\vec k=\vec k_1\mid\mid\vec  k_2$ consists of two $N$-bit keys, and hence is of length $2N$.
For an encrypted message $\mathsf{EM}_{\vec k_1,\vec k_2}(\vec x)=\vec y$, the original message $\vec x$ can be recovered using the same secret keys and the inverse permutation $P^{-1}$ by reversing the order of operations $\vec x = P^{-1}(\vec y\oplus \vec k_2)\oplus \vec k_1$.
The schema of the cipher is illustrated in Fig.~\ref{fig:cryptographic-primitives}(a).

Classically, the attack model considers an adversary with access to an encryption oracle in addition to the permutation $P$, which is publicly available, and gives a tight bound on the security of the scheme of $\Theta(2^N)$ \cite{emrevisited1}. 

\subsection{Simon's Algorithm for the Even-Mansour Cipher}

A polynomial-time key recovery attack showed that the single-round, two-key version of the Even-Mansour cipher is breakable in a quantum setting \cite{kuwakado}.
The quantum model assumes that the cipher is accessible as an encryption oracle operating on quantum information.
Given an $N$-qubit state $\vec x$ as plaintext, the quantum gate
\begin{eqnarray*}
U_{\mathsf{EM}_{\vec k_1,\vec k_2}}\ket{\vec x}\ket{\vec a}=\ket{\vec x}\ket{\vec a\oplus\mathsf{EM}_{\vec k_1,\vec k_2}(\vec x)},
\end{eqnarray*}
presented in Fig.~\ref{fig:cryptographic-primitives}(b), creates an $N$-qubit state that represents the Even-Mansour encryption of $\vec x$. 
The aim is to recover $\vec k_1$ and $\vec k_2$ which would require $\mathcal{O}(2^{N/2})$ time using Grover's search algorithm. 
The attack uses the structure of the Even-Mansour cipher to define a two-to-one function with period $\vec k_1$ for which Simon's algorithm recovers the period, and hence the key, in linear time.

For a given cipher $\mathsf{EM}_{\vec k_1,\vec k_2}$ with permutation $P$, the idea is to consider the function
\begin{eqnarray}\label{eqn:EM_function}
f(\vec x)=\mathsf{EM}_{\vec k_1,\vec k_2}(\vec x)\oplus P(\vec x)
\end{eqnarray}
(with the corresponding query gate $U_f$) which has period $\vec k_1$. 
For a random permutation $P$, it is unlikely that there is an additional collision in $\vec x$ and $\vec y$ with $\vec y\neq \vec x\oplus \vec k_1$ and $f(\vec x)=f(\vec y)$.
Therefore, with high probability, $f$ is a two-to-one function and Simon's algorithm can be applied to recover the period.

Simon's algorithm will yield a sequence $\vec y_0,\vec y_1, ...$ with $\vec y_i \cdot \vec k_1=0$, which can be solved for $\vec k_1$. 
Using the oracle $\mathsf{EM}_{\vec k_1, \vec k_2}$, the public permutation and an arbitrarily chosen message $\vec m\in\{0,1\}^N$, we can compute 
\begin{eqnarray*}
\mathsf{EM}_{\vec k_1,\vec k_2}(\vec m)\oplus P(\vec m\oplus \vec k_1) =\vec k_2
\end{eqnarray*}
to recover the second part of the key $\vec k_2$.

The quantum complexity is dominated by obtaining at least $N-1$ equations and results in running $\mathcal{O}(N)$ iterations of Simon's circuit, and therefore requires $\mathcal{O}(N)$ quantum queries, and $\mathcal{O}(N^3)$ classical operations in the post-processing step.
In terms of running time, this provides an exponential speedup compared to the classical security bound of Even-Mansour.

Note that the above attack assumes access to an encryption oracle on quantum superposition states. In practice, this is of course not a realistic setting. However, the attack proves that, in principle, exponential speedups compared to the best classical attack are possible for symmetric ciphers. Furthermore,  the assumption of an encryption oracle on superposition states can be avoided at the cost of increasing the complexity of the attack \cite{em_hosoyamada}.

\begin{figure*}[ht]
\centering
\includegraphics[width=1.0\linewidth]{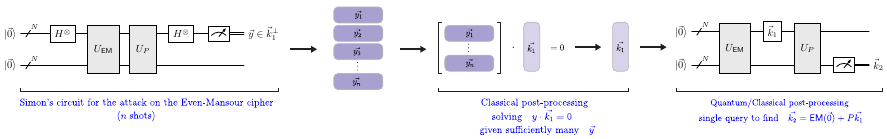}
\caption{Schematic workflow of the attack: Quantum sampling via Simon’s algorithm and final quantum/classical secret key extraction.}
 \label{fig:attack-workflow}
\end{figure*}

\section{Data and Experiments}
\label{sec:experiments}
\subsection{Substitution Boxes (S-boxes)}
In the Even-Mansour symmetric block cipher construction, the core cryptographic mixing relies on a single, publicly known, non-linear permutation $P$. For our proof of concept implementation, for $N=3,4$ we take $P$ to be a bijective map whose construction is inspired by the AES substitution box (S-box) \cite{AES}. This allows for a better comparison between different sizes of bitstrings. For this, we view $\{0,1\}^N$ as the finite field $\mathbb{F}_{2^N}$ obtained by taking the quotient of the polynomial ring $\mathbb{F}_2[X]$ with respect to a suitable irreducible polynomial of degree $N$. Our permutation on $\{0,1\}^N$ is then given by the inversion $a\longmapsto a^{-1}$ in  $\mathbb{F}_{2^N}$ (with $0$ mapping to $0$ by definition) followed by a suitable affine transformation of the $N$-dimensional vector space $\mathbb{F}_{2^N}$ over $\mathbb{F}_2$. 

\begin{figure*}[b]
    \centering
    \includegraphics[width=0.9\linewidth]{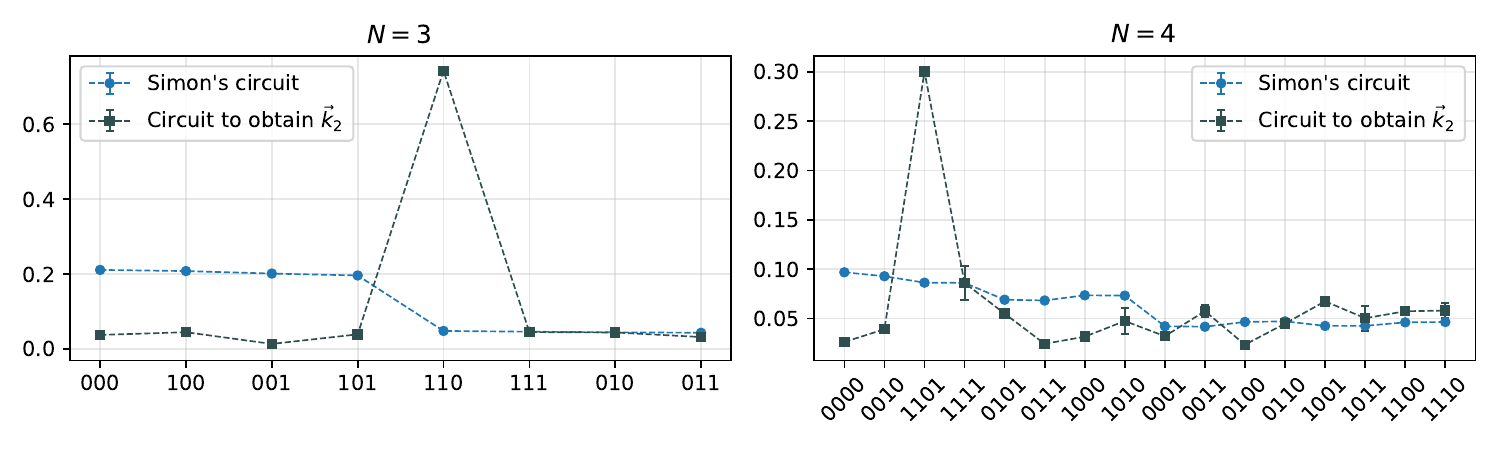}
    \caption{Mean relative distribution of five independent 3 and 4-bit trials on ibm\_miami. Small deviations demonstrate hardware reproducibility, with all individual runs succeeding.}
    \label{fig:results_3_4}
\end{figure*}

We do not claim that our permutations are secure cryptographic S-boxes. For $N=3,4$ the permutations are listed explicitly in section \ref{sec:results}. To execute Simon's period-finding algorithm, these classical lookup tables must be synthesized into fully reversible quantum circuits, which include fundamental quantum operations such as Pauli-X, Toffoli, and SWAP gates, see Sec.~\ref{Quantum Circuit Synthesis and the DORCIS Toolchain}. Unfortunately, as explained therein, we only managed to construct quantum circuits for the permutations in the cases $N=3,4$. 

\subsection{Quantum Circuit Synthesis and the DORCIS Toolchain}
\label{Quantum Circuit Synthesis and the DORCIS Toolchain}
To implement the Even-Mansour encryption oracle on physical hardware, the classical non-linear public permutation (the S-box) must first be translated into a fully reversible quantum circuit. For this critical synthesis step, we employed the Depth Optimized Quantum Implementation of Substitution Boxes (DORCIS) toolchain \cite{dorcis}. DORCIS takes the classical permutation mapping tables and algorithmically decomposes them into an optimized sequence of elementary quantum operations, specifically aiming to minimize the overall circuit depth. For our experiments, this decomposition primarily yielded a combination of Pauli-X (NOT), Toffoli (CCNOT), and SWAP gates. DORCIS then tries to optimize the depth of these permutation circuits to yield an optimal unitary for the hardware.

Despite its effectiveness for our small-scale proof of concept experiments, our reliance on this classical synthesis toolchain revealed a scaling limitation. While the quantum hardware itself is steadily maturing, the classical pre-processing required to optimize these quantum circuits encounters a critical memory bottleneck. Specifically, adapting the DORCIS algorithm for key lengths of $N=5$ and beyond failed to generate quantum circuits. DORCIS was executed on a dual socket Intel(R) Xeon(R) Platinum 8480+ 56 core CPU (amounting to 224 logical compute nodes with multithreading enabled) equipped with 3.9 TiB of memory. This classical computational failure prevented the experimental evaluation of larger cryptographic instances.
It should be clarified that DORCIS was primarily designed for the depth-optimized implementation of small-scale S-boxes, however it can still be interesting for future work to explore circuit synthetization for larger and more general permutations for Even-Mansour ciphers.

\subsection{Quantum Hardware and Software Framework}
To practically evaluate the Even-Mansour cipher, our quantum circuits were implemented and executed on the ibm\_miami processor using Qiskit for circuit design and hardware-optimized transpilation \cite{qiskit2024}. This optimization was crucial for minimizing SWAP gates and reducing overall circuit depth. We specifically leveraged ibm\_miami's new lattice qubit topology. We bypassed automatic routing and explicitly mapped our logical circuits to continuous physical qubit subgraphs: specifically, we selected physical qubits [0, 1, 2, 12, 11, 10] for the $N=3$ key length experiments, and [0, 1, 2, 3, 13, 12, 11, 10] for the $N=4$ key length experiments.

\subsection{Error Mitigation Techniques}
To combat decoherence during the extended idle periods inherent to the deep, multi-controlled structures of the Even-Mansour S-boxes, we implemented Dynamical Decoupling (DD) \cite{DD1, DD2}. In the transpilation pipeline, DD was applied by scheduling symmetric sequences of unitaries (such as alternating X pulses) on idle physical qubits. This technique effectively refocuses the quantum states, suppressing low-frequency environmental noise and mitigating crosstalk that can occur within the densely connected lattice topology of ibm\_miami.

Furthermore, we addressed the accumulation of coherent errors generated by the extensive number of two-qubit gates required for the algorithm by employing Pauli Twirling \cite{PT1, PT2}. This technique mitigates worst-case coherent noise by tailoring it into a more predictable, stochastic Pauli channel. This was achieved by systematically inserting randomly selected, but logically equivalent, pairs of Pauli operators before and after the targeted gates throughout the transpiled circuit. By averaging the measurement outcomes over multiple twirled variations of the algorithmic circuit, we in theory prevent the constructive interference of coherent gate errors. This is particularly critical for period-finding algorithms, where systematic phase distortions can otherwise obscure the algorithmically structural collisions necessary to extract the hidden key.

\section{Results and Discussion}
\label{sec:results}

Each experiment is split into two steps. In the first step, we use a quantum circuit for Simon's algorithm to obtain a distribution of vectors. The most commonly appearing results are interpreted as the sequence of vectors from which we can compute the first secret key. 
In the second step, we use the obtained first secret key and the quantum oracle to compute the second key. It should be noted that this step can either be executed on quantum hardware or classically. 
An overview of the attack workflow is presented in Fig.~\ref{fig:attack-workflow}.

\subsection{3-bit Case}
 
\begin{figure}[!h]
\centering
\begin{quantikz}
\lstick{$\ket{x_0}$} & \ctrl{2} & \targ{} & \ctrl{2} & & \targ{} & \swap{1} & \rstick{$\ket{y_0}$} \\
\lstick{$\ket{x_1}$} & \control{} & & \targ{} & \targ{} & \ctrl{-1} & \targX{} & \rstick{$\ket{y_1}$} \\
\lstick{$\ket{x_2}$} & \targ{} & & \control{} & \targ{} & \ctrl{-1} & & \rstick{$\ket{y_2}$}
\end{quantikz}
    \caption{Quantum circuit for the 3-qubit permutation.}
    \label{fig:permutation3}
\end{figure}
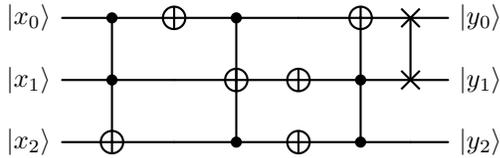

For $N=3$, the constructed permutation $P$ in look-up table (LUT) format is \texttt{52367401}.
Using DORCIS, we obtained the quantum circuit implementation for the permutation which is depicted in Fig.~\ref{fig:permutation3}.
It has three qubits $(x_0, x_1, x_2)$ and three different kinds of quantum gates acting on them.
There are three NOT gates, three Toffoli (CCNOT) gates, each targeting a different qubit, denoted as $\bullet$ and $\oplus$ for their control and target qubits, respectively, and a SWAP gate.
The {\it depth} of a circuit is the number of gates, counting any set of gates that can run in parallel as a single gate.
The {\it T-depth} is the number of non-Clifford gates, again counting any set of non-Clifford gates that can run in parallel as a single gate.
Because the decomposition of a Toffoli gate has depth $7$, the circuit for the permutation has depth 23 and T-depth 3.

Given the quantum oracle, after running 5 independent experiments, each with $10^5$ iterations of Simon's algorithm, we obtain the results listed on the left side in Table~\ref{tab:results3}. 
Fig.~\ref{fig:results_3_4} shows the results distribution.

\begin{table}
\centering
\caption{Mean results for the 3-bit case of 5 independent experiments on ibm\_miami with $10^5$ shots each.}
  \footnotesize
    \begin{tabular}{lr|lr}
      \toprule
      \multicolumn{2}{c}{\textbf{Simon's circuit}} & \multicolumn{2}{c}{\textbf{Circuit to obtain $\vec k_2$}} \\
      Result & Counts & Result & Counts\\
      \midrule
      \texttt{000} & $21\,118.2$ & \texttt{000} & $3\,772.8$ \\
      \texttt{001} & $20\,142.8$ & \texttt{001} & $1\,373.8$ \\
      \texttt{010} & $4\,473.2$ & \texttt{010} & $4\,427.6$ \\
      \texttt{011} & $4\,354$ & \texttt{011} & $3\,210.8$ \\
      \texttt{100} & $20\,808.8$ & \texttt{100} & $4\,502.8$ \\
      \texttt{101} & $19\,637$ & \texttt{101} & $3\,928.8$ \\
      \texttt{110} & $4\,836.6$ & \texttt{110} & $74\,252.2$ \\
      \texttt{111} & $4\,629.4$ & \texttt{111} & $4\,531.2$ \\
      \bottomrule
    \end{tabular}
  \label{tab:results3}
\end{table}

For any vector $\vec v\in \{0,1\}^3$, there are $2^{2}=4$ vectors orthogonal to $\vec v$.
We therefore expect a distribution over the four results that are orthogonal to $\vec k_1$.
The four most common results are $\{000, 100,001,101\}$.
Since the vector space $\vec{k_1}^\bot$ has dimension $2$, by picking two linearly independent vectors $100$ and $001$ and solving
\begin{eqnarray*}
\begin{bmatrix}  1 & 0 & 0 \\  0 & 0 & 1 \end{bmatrix} \cdot \left[ \begin{array}{c} k_{1,0} \\ k_{1,1} \\ k_{1,2} \end{array} \right] = \left[ \begin{array}{c} 0 \\ 0 \end{array} \right],
\end{eqnarray*}
we obtain $\vec k_1=(0,1,0)$.

To obtain the second part of the key, we compute $\vec k_2 = \mathsf{EM}_{\vec k_1,\vec k_2}(\vec 0)\oplus P(\vec k_1)$, where the encryption $\mathsf{EM}$ of the message $000$, can be retrieved by a call to the encryption oracle. 
The results are also shown in Table~\ref{tab:results3} and Fig.~\ref{fig:results_3_4} and the most common result is $\vec k_2=(1,1,0)$. All of the 5 experiments were also successful on their own indicating reproducibility. 
With this result we successfully implemented Simon's attack on the Even-Mansour cipher.

\subsection{4-bit Case}

\begin{figure}[ht]
\centering
\resizebox{1\linewidth}{!}{
\begin{quantikz}
\lstick{$\ket{x_0}$}&\targ{}&\ctrl{2}&\targ{}&&\ctrl{3}&\ctrl{3}&\ctrl{1}&\targ{}&&\targ{}&\ctrl{2}&\targ{}&\swap{1}&\rstick{$\ket{y_0}$}\\
\lstick{$\ket{x_1}$}&&\control{}&&\targ{}&\targ{}&\control{}&\targ{}&\control{}&\targ{}&&\control{}&&\targX{}&\rstick{$\ket{y_1}$}\\
\lstick{$\ket{x_2}$}&&\targ{}&\ctrl{-2}&\targ{}&&\control{}&\targ{}&&\ctrl{-1}&&\targ{}&\ctrl{-2}&\swap{1}&\rstick{$\ket{y_2}$}\\
\lstick{$\ket{x_3}$}&\ctrl{-3}&&&&\control{}&\targ{}&\ctrl{-1}&\ctrl{-3}&&\ctrl{-3}&&&\targX{}&\rstick{$\ket{y_3}$}
\end{quantikz}
}
    \caption{Quantum circuit for the 4-qubit permutation.}
    \label{fig:permutation4}
\end{figure}
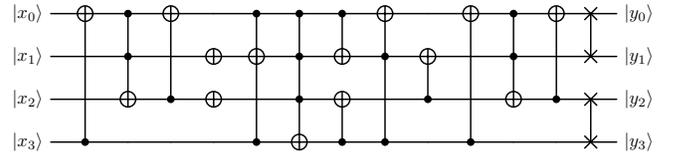

For $N=4$, the constructed permutation $P$ in LUT format is \texttt{E4B238091A7F6C5D}.
The quantum circuit obtained using DORCIS is depicted in Fig.~\ref{fig:permutation4}.
It has four qubits $(x_0, x_1, x_2, x_4)$ and the circuit has depth 54 and T-depth 7.

Given the quantum oracle, after running 5 independent experiments, each with $10^5$ shots, we obtain the results listed on the left side in Table~\ref{tab:results4}. 
Fig.~\ref{fig:results_3_4} shows the results distribution.

\begin{table*}
\centering
  \caption{Mean results for the 4-bit case of 5 independent experiments on ibm\_miami with $10^5$ shots each.}
  \footnotesize
    \begin{tabular}{lrlr|lrlr}
      \toprule
      \multicolumn{4}{c}{\textbf{Simon's circuit}} & \multicolumn{4}{c}{\textbf{Circuit to obtain $\vec k_2$}} \\
      Result & Counts & Result & Counts & Result & Counts & Result & Counts\\
      \midrule
      \texttt{0000} & $9\,676.8$ & \texttt{1000} & $7\,345.4$ & \texttt{0000} & $2\,613.8$ & \texttt{1000} & $3\,149.4$ \\
      \texttt{0001} & $4\,203.6$ & \texttt{1001} & $4\,241.6$ & \texttt{0001} & $3\,198.8$ & \texttt{1001} & $6\,737$ \\
      \texttt{0010} & $9\,277.8$ & \texttt{1010} & $7\,311.2$ & \texttt{0010} & $3\,903.4$ & \texttt{1010} & $4\,739$ \\
      \texttt{0011} & $4\,161.6$ & \texttt{1011} & $4\,250.2$ & \texttt{0011} & $5\,738.4$ & \texttt{1011} & $5\,020$ \\
      \texttt{0100} & $4\,642.8$ & \texttt{1100} & $4\,602.6$ & \texttt{0100} & $2\,333.2$ & \texttt{1100} & $5\,728.4$ \\
      \texttt{0101} & $6\,895.8$ & \texttt{1101} & $8\,626.2$ & \texttt{0101} & $5\,492.4$ & \texttt{1101} & $30\,036.8$ \\
      \texttt{0110} & $4\,705.6$ & \texttt{1110} & $4\,635$ & \texttt{0110} & $2\,425.6$ & \texttt{1110} & $5\,797.6$ \\
      \texttt{0111} & $6\,819.4$ & \texttt{1111} & $8\,604.4$ & \texttt{0111} & $4\,484$ & \texttt{1111} & $8\,602.2$ \\
      \bottomrule
    \end{tabular}
  \label{tab:results4}
\end{table*}

For any vector $\vec v\in \{0,1\}^4$, there are $2^{3}=8$ vectors orthogonal to $\vec v$, so we expect a distribution over eight results.
The eight most commonly occurring results in descending order are $\{0000,0010,1101,1111,0101,0111,1000,1010\}$.
Since the vector space $\vec{k_1}^\bot$ has dimension $3$, by picking the three vectors $0010$, $1101$, $1010$ and solving the resulting system of linear equations, we obtain $\vec k_1=0101$.

To obtain the second part of the key, we compute $\vec k_2 = \mathsf{EM}_{\vec k_1,\vec k_2}(\vec 0)\oplus P(\vec k_1)$. 
The corresponding results are also shown in Table~\ref{tab:results4} and Fig.~\ref{fig:results_3_4} and the most common result is $\vec k_2=(1,1,0,1)$. 
All of the 5 experiments were also successful on their own indicating reproducibility.
Again, with this results the attack was successful. 

\subsection{\label{section:error_analysis}Qualitative Error Analysis}

\textit{\textbf{General Error Sources.}}
Simon's algorithm on real quantum hardware is naturally subjected to errors originating both from intrinsic design and hardware noise. 

As soon as short keys are chosen, random collisions with unwanted keys that do not originate from the periodicity of the Even-Mansour cipher with its generating key need to be considered.
The investigation of an imperfect Simon's promise for symmetric block ciphers subjected to local, random collisions showed that this obstacle vanishes naturally as the key lengths increase \cite{kaplan}.
If $\varepsilon <1$ denotes the probability that the image of the function $f$ defined in Eq. (\ref{eqn:EM_function}) matches a different key than $\vec k_1$, i.e.
\begin{equation*}\label{eq:error_Simons_promise}
    \varepsilon\equiv \varepsilon(f,\vec k_1):= \max_{t\in\{0,1\}^N \setminus \{\vec 0,\vec k_1\}} \mathbb{P}_{\vec x}\left[f(\vec x) = f(\vec x \oplus\vec t) \right],
\end{equation*}
where the probability distribution of $\vec{x}$ is taken to be uniform on $\{0,1\}^N$, the success probability of Simon's algorithm after $cN$ queries, is $P_{\text{succ}} = 1 - \left(2\left( \frac{1+\varepsilon}{2}\right)^c \right)^N$ \cite{kaplan}.
For random functions, $\varepsilon\propto\Theta(N 2^{-N})$, whereas for constant functions, no period recovery is possible at all $(\varepsilon=1)$. 
If Simon's algorithm is realized for $3$--$5$ bit keys via low-shot quantum computations ($cN \sim 10^1, 10^2$), $\varepsilon$ represents a relevant error source. For our chosen 4-bit S-box, the random collision error reaches $\varepsilon= 0.25$, whereas for the 3-bit S-box, $\varepsilon=0$. 
Since we perform $cN = 10^5$ shots, we can neglect local collisions and explain the experimental deviation solely by hardware errors.

\begin{figure}
    \centering
    \includegraphics[width=\linewidth]{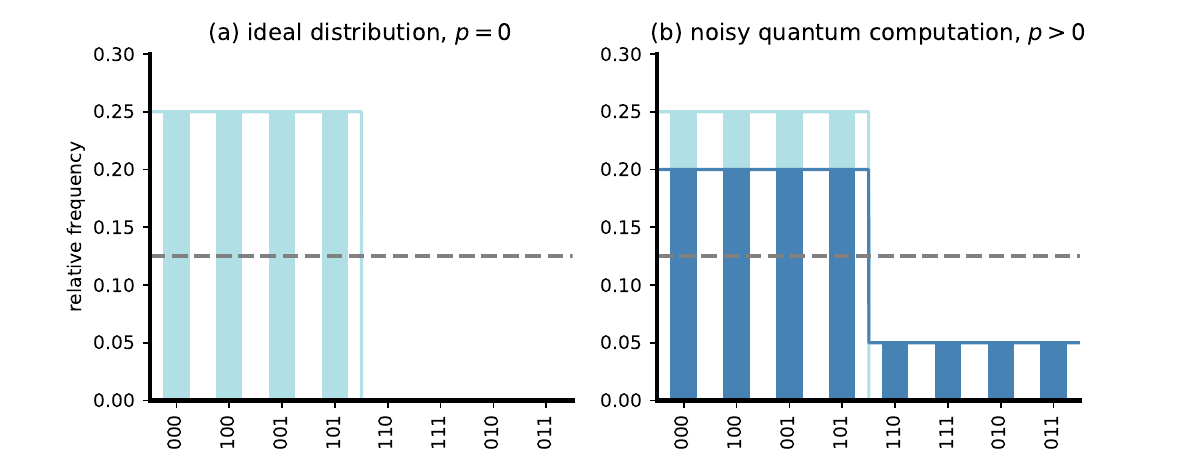}
    \caption{Schematic result of Simon's circuit for the 3-bit case. (a) No noise for $p=0$, (b) if $p>0$, depolarizing noise pulls towards a uniform distribution.}
    \label{fig:depol_noise}
\end{figure}

\textit{\textbf{Hardware Errors.}}
There are three dominating error sources affecting noisy-intermediate scale quantum computation \cite{georgopoulos2021modeling}: depolarizing noise, state preparation and measurement (SPAM) errors, and decoherence. We mainly focus qualitatively on the symmetric depolarization channel, which encapsulates hardware infidelities \cite{georgopoulos2021modeling}. 
Its closed form for a $d-$dimensional multi-qubit state $\rho$, reads $D_p(\rho) = (1-p)\rho +\frac{p}{d} I$.
Let $\mathbb{b}:= \{0,1\}$ and $\mathbb{b}^N$ be the vector space of $N$-bit strings. 
The ideal result of Simon's algorithm provides a uniform distribution over one half of $\mathbb{b}^N$.  
This is shown in Fig.~\ref{fig:depol_noise} and to be compared with Fig.~\ref{fig:results_3_4}. 
The depolarization noise randomizes the ideal result towards a uniform distribution on $\mathbb{b}^N$, as follows. For an $N$-bit key, we denote the subset resulting from Simon's circuit by $S\subset \mathbb{b}^N$ and its complement by $\overline{S}$. For our 3-bit computation, $S = \{000, 100, 001, 101 \}$. We employ a simplified, generic noise treatment through symmetric depolarization \cite{nielsen}, which estimates the resulting distribution $\sigma$ with a noise parameter $p\in[0,1]$ via
\begin{align*}\label{eq:depolarization_distribution}
    \sigma_p(\vec{x}) = \begin{cases}
        \left(1-\frac{p}{2}\right) \frac{1}{2^{N-1}}, &\text{ if }\vec{x}\in S\\
         \frac{p}{2^N}, &\text{ if }\vec{x}\in\overline{S}.
    \end{cases}
\end{align*}
To estimate an effective value for $p$, one computes a naive estimation on the error rate of hardware $p \approx M p_{gate}$, where $p_{gate} \approx 10^{-3}$ denotes the averaged gate error and $M = 434$ denotes the number of total pulses used on the superconducting hardware, required for the attack. Using this naive error model, we get $p\approx0.434$, so $\sigma_p$ roughly matches the distribution observed in Fig.~\ref{fig:results_3_4}.

\section{Conclusion}
\label{sec:conclusion}
In this work, we presented a practical proof of concept for executing a quantum key recovery attack against the Even-Mansour cipher using Simon's algorithm. By deploying error mitigated quantum circuits on the IBM quantum processor ibm\_miami, we successfully recovered the secret keys for 3-bit and 4-bit S-boxes. These results validate the theoretical exponential speedups offered by quantum period-finding algorithms, confirming the feasibility of structural exploitations of the Even-Mansour scheme in a quantum setting.

However, our investigation also exposed a limitation in the classical pre-processing pipeline.
The classical toolchain required for the quantum circuit depth optimization of the underlying S-boxes encountered a memory bottleneck. Specifically, the classical synthesis failed to produce optimized quantum circuits for the public permutations at $N=5$.
For future work we propose the investigation of other optimized circuit synthetization tools for larger permutations, possibly utilizing heuristic or classical machine learning algorithms.

\clearpage
\bibliographystyle{IEEEtran}
\bibliography{references.bib}

\end{document}